\documentclass[fleqn,usenatbib]{mnras}

\usepackage[T1]{fontenc}

\usepackage{graphicx}	
\usepackage{amsmath}	
\usepackage{amssymb}	
\usepackage{lscape}
\usepackage{ulem}
\usepackage{tabularx}
\usepackage[dvipsnames]{xcolor}
\usepackage{soul}
\usepackage{caption}
\usepackage{subcaption}
\RequirePackage{tikz}

\graphicspath{{images/}}
	
	\title[Circularly symmetry extended diffuse radio emission]{Discovery of a circularly symmetric extended diffuse radio emission
around an elliptical galaxy with the VLA FIRST survey
}

\author[S. Kumari and S. Pal]{
Shobha Kumari$^{1}$\orcidA{}
Sabyasachi Pal$^{1}$\orcidB\thanks{E-mail: sabya.pal@gmail.com (SP)}\\
$^{1}$Midnapore City College, Kuturia, Bhadutala, Paschim Medinipur, West Bengal, 721129, India \\
}

\newcommand{\orcidicon}{\includegraphics[width=0.32cm]{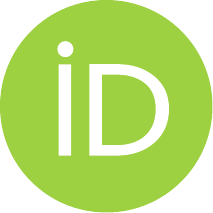}}

\foreach \x in {A, ..., Z}{%
\expandafter\xdef\csname orcid\x\endcsname{\noexpand\href{https://orcid.org/\csname orcidauthor\x\endcsname}{\noexpand\orcidicon}}
}

\date{Accepted 2023 December 20. Received 2023 December 18; in original form 2023 August 22}

\pubyear{2024}
 \usepackage{newtxtext,newtxmath}
\begin{document}
\label{firstpage}
\pagerange{\pageref{firstpage}--\pageref{lastpage}}
	\maketitle

\def\src{J1507+3013}
\begin{abstract}
{We identify a source (J1507+3013) with an extended diffuse radio emission around an elliptical galaxy from the Very Large Array (VLA) Faint Images of Radio Sky at Twenty-cm (FIRST) survey. J1507+3013 possesses a morphology similar to the recently identified circular, low-surface-brightness, edge-brightened radio sources commonly known as odd radio circles (ORCs). Such diffuse emissions, as reported in this paper, are also found in mini-haloes and fossil radio galaxies, but the results presented here do not match the properties of mini-haloes or of fossil radio galaxies. The extended emission observed in J1507+3013 around an elliptical galaxy is a very rare class of diffuse emission that is unlike any previously known class of diffuse emission. The extended diffuse emission of J1507+3013 is also detected in the Low Frequency Array (LOFAR) at 144 MHz. J1507+3013 is hosted by an optical galaxy near the geometrical centre of the structure with a photometric redshift of $z=0.079$. The physical extent of J1507+3013 is approximately 68 kpc, with a peak-to-peak angular size of 44 arcsec. It shows significantly higher flux
densities compared with previously discovered ORCs. The spectral index of J1507+3013 varies between --0.90 and --1.4 in different regions of the diffuse structure, which is comparable to the case for previously discovered ORCs but less steep than for mini-haloes and fossil radio galaxies. If we consider J1507+3013 as a candidate ORC, then this would be the closest and most luminous ORC discovered so far. This paper describes the radio, spectral, and optical/IR properties of J1507+3013 in order to study the nature of this source.} 
\end{abstract}




\section{Introduction}
\label{sec:intro}
Radio galaxies exhibit a broad range of morphological structures. A compact radio core coincides with the optical nucleus detected in most radio galaxies. The interaction of energetic jets with hot diffuse gas and the surrounding environment of the host galaxy determines the structure of a typical radio galaxy. Jets are composed of radiative plasma consisting of electrons, positrons, and/or protons. Jets of radio galaxies can extend into the intergalactic medium (IGM) on scales ranging from parsecs to hundreds of kiloparsecs. When the jet collides with the interstellar medium (ISM), it is compelled to recollimate, which might cause internal shocks, leading to slowing down jets and facilitating the mixing of the surrounding material \citep{Pe07}. The jet may also be slowed down by individual obstacles. For example, the winds of a single asymptotic giant branch star \citep{Pe17} or a supernova explosion occurring inside the jet and filling the jet with its remnant \citep{Vi17} may be sufficient to impact the dynamics of the structure of radio sources. Supermassive black holes (SMBHs) within a mass range of 10$^6-10^{9.5}$ M$_{\odot}$ reside at the centre of bulges and elliptical galaxies, and feedback from these SMBHs can also profoundly affect the formation and evolution of the morphology of radio sources \citep{Ko95, Si98, Ki03, Gr04, Sp05, Ho06}.    

Diffuse radio emission with a diameter of an arcminute or more around an elliptical galaxy (e.g. Fig. 8a, b of \citet{Sa02}) having no association with any galaxy cluster is very rarely observed. In the search for such diffuse emission around an elliptical galaxy using the Very Large Array (VLA) Faint Images of Radio Sky at Twenty-cm (FIRST) survey, \citet{Ku23} recently identified J1407+0453, a source with a horseshoe-shaped inner ring of diffuse emission. The physical extent of the total diffuse emission of J1407+0453 is measured as $\sim$160 kpc, whereas the physical extent of the horseshoe-shaped inner ring is $\sim$ 25 kpc. 
Galaxy clusters often have diffuse radio emissions, with sizes ranging from kiloparsecs to megaparsecs. These diffuse emissions are commonly
found in radio haloes, radio relics, and mini-haloes \citep{Kem04, Fe21}. Their morphology is often asymmetrical, superimposed on constituent galaxy radio emission, and often with a diffused remnant towards the edge. Some diffuse emissions have been observed as remnants of central active galactic nucleus (AGN) activity in the distant past and are typically known as fossil radio galaxies \citep{Kem04, Ri23}. The AGN in
a fossil radio galaxy no longer actively accretes material, but its past impact on galaxies is preserved in the form of diffuse radio structures.

Recently, odd radio circles (ORCs), which are a very new class of object with diffuse, edge-brightened radio emissions, have been identified \citep{No21a, No21b, No21c, Ko21, Om22}. These diffuse sources exhibit a ring-like morphology of about an arcminute in diameter. Astronomers have conflicting views regarding the nature of ORCs. Some of the possible mechanisms proposed to explain the ring-like structures in ORCs include supernova remnants (SNRs), cluster haloes, end-on lobes of radio galaxies, shockwaves from the collision of cosmic structures, emissions from the spherical remnants of energetic radio galaxies, and the effect of dark matter decay \citep{No21a, No21b, No21c, Ko21, No22}. 

Presently, six sources [ORC J2103--6200 (ORC1), ORC J2058-5736 (ORC2), ORC J2059-5736 (ORC3), ORC J1555+2726 (ORC4), ORC J0102--2450 (ORC5) and ORC J0020+3018 (ORC6)] have been identified with morphologies corresponding to those of the ORCs. Three of the six ORCs are known to contain optical galaxies at their geometrical center. 
A negative spectral index (assuming, $S_{\nu} \propto \nu^{\alpha}$) in the radio emission from ORCs is observed, which indicates that the radio emission is non-thermal and most likely originates as a result of synchrotron radiation.

\begin{figure}
\vbox{
\centerline{
\includegraphics[width=9.1cm, height=8.0cm, origin=c]{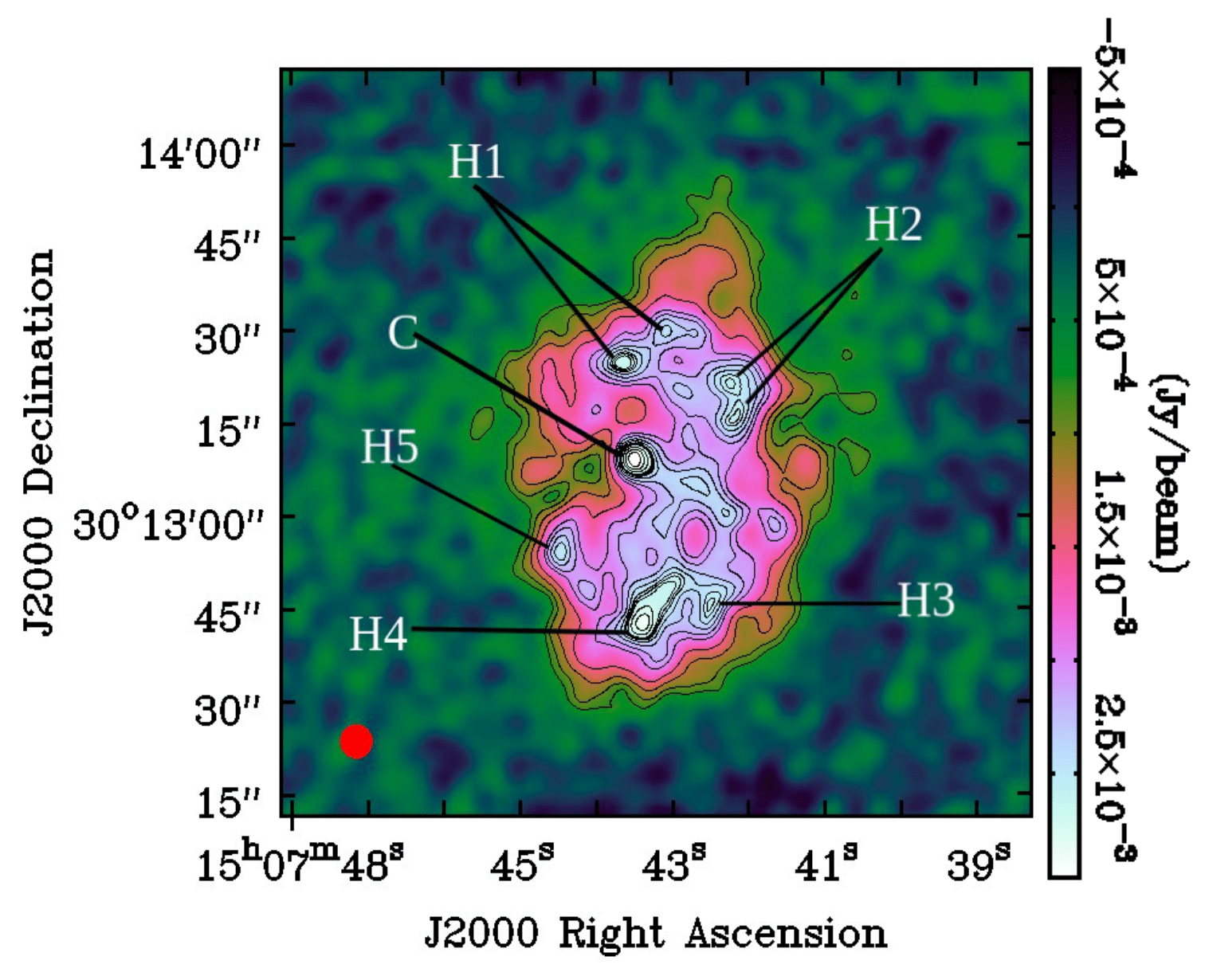}
}
}
\caption{ J1507+3013 is identified at 1400 MHz in the VLA FIRST survey. The contour levels are at 3$\sigma\times$(1, 1.4, 2, 2.8, 4, 4.5, 5, 5.4, 5.7, 5.8, 5.9, 6, 6.2, 6.5, 7, 7.2, 7.4), where $\sigma=133$  $\mu$Jy beam$^{-1}$ is the rms noise of the FIRST map. The colour scale represents the flux density.}
\label{fig:ORC_rad}
\end{figure}

This paper is organized as follows: Source identification is described in Section \ref{sec:identi}. In Section \ref{sec:properties}, we elaborate on the properties of the newly discovered source. In Section \ref{sec:result}, we discuss our results. Section \ref{sec:conclusion} summarizes the results of this study.
We use the following $\Lambda$CDM cosmology parameters in this paper from the final full-mission {\it Planck} measurements of the cosmic microwave background anisotropies: $H_0$ = 67.4 km s$^{-1}$ Mpc$^{-1}$, $\Omega_m$ = 0.315, and $\Omega_{vac}$ = 0.685  \citep{Ag20}.

\section{J1507+3013: A circularly symmetry extended diffuse radio emission around an elliptical galaxy}
\label{sec:identi}
\subsection{J1507+3013 in the 1400-MHz VLA FIRST survey}
\label{subsec:vla}
\begin{figure}
\vbox{
\centerline{
\includegraphics[width=8.8cm,origin=c]{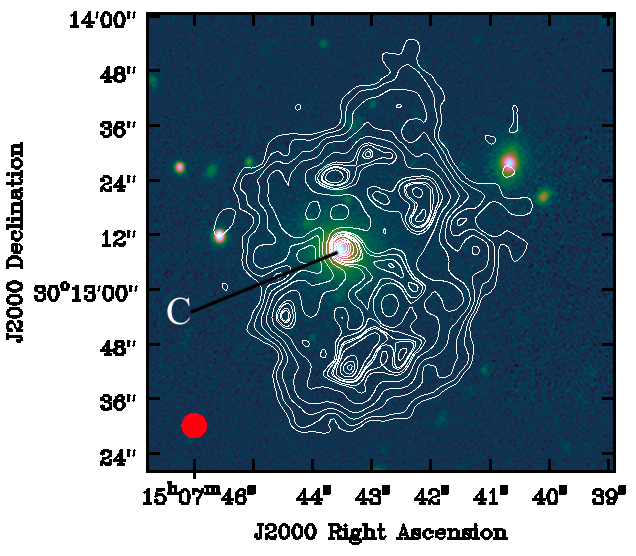}
}
}
\caption{J1507+3013 at 1400 MHz with the Pan-STARRS optical r band \citep{Fl20} superimposed. The contour levels are at 3$\sigma\times$(1, 1.4, 2, 2.8, 4, 4.5, 5, 5.4, 5.7, 5.8, 5.9, 6, 6.2, 6.5, 7, 7.2, 7.4), where $\sigma=133$  $\mu$Jy beam$^{-1}$ is the rms noise of the FIRST map.}
\label{fig:ORC_rad_op}
\end{figure}

We identify a source (J1507+3013) with an extended diffuse radio emission around an elliptical galaxy with nearly circular symmetry from the Very Large Array (VLA) Faint Images of Radio Sky at Twenty-cm (FIRST) surv e y at 1400 MHz (see the VLA FIRST image of J1507+3013 in Figs. \ref{fig:ORC_rad} and \ref{fig:ORC_rad_op}) \citep{Be95, Wh97}.
J1507+3013 is identified accidentally during the search for hybrid morphology radio sources (HyMoRS; \citet{Pa22}) from the VLA FIRST survey. The FIRST catalogue contain 946432 radio sources. The selection criteria are stated in \citet{Pa22}, where we filtered the FIRST catalogue sources with an angular size of $\geq15\arcsec$ (i.e., three times the convolution beam size). As a result, our filtering produced a total of 20045 sources. To search for radio galaxies with rare morphologies, we visually examined the field of all the 20045 filtered sources from the FIRST survey.

\begin{figure}
\includegraphics[width=9cm, origin=c]{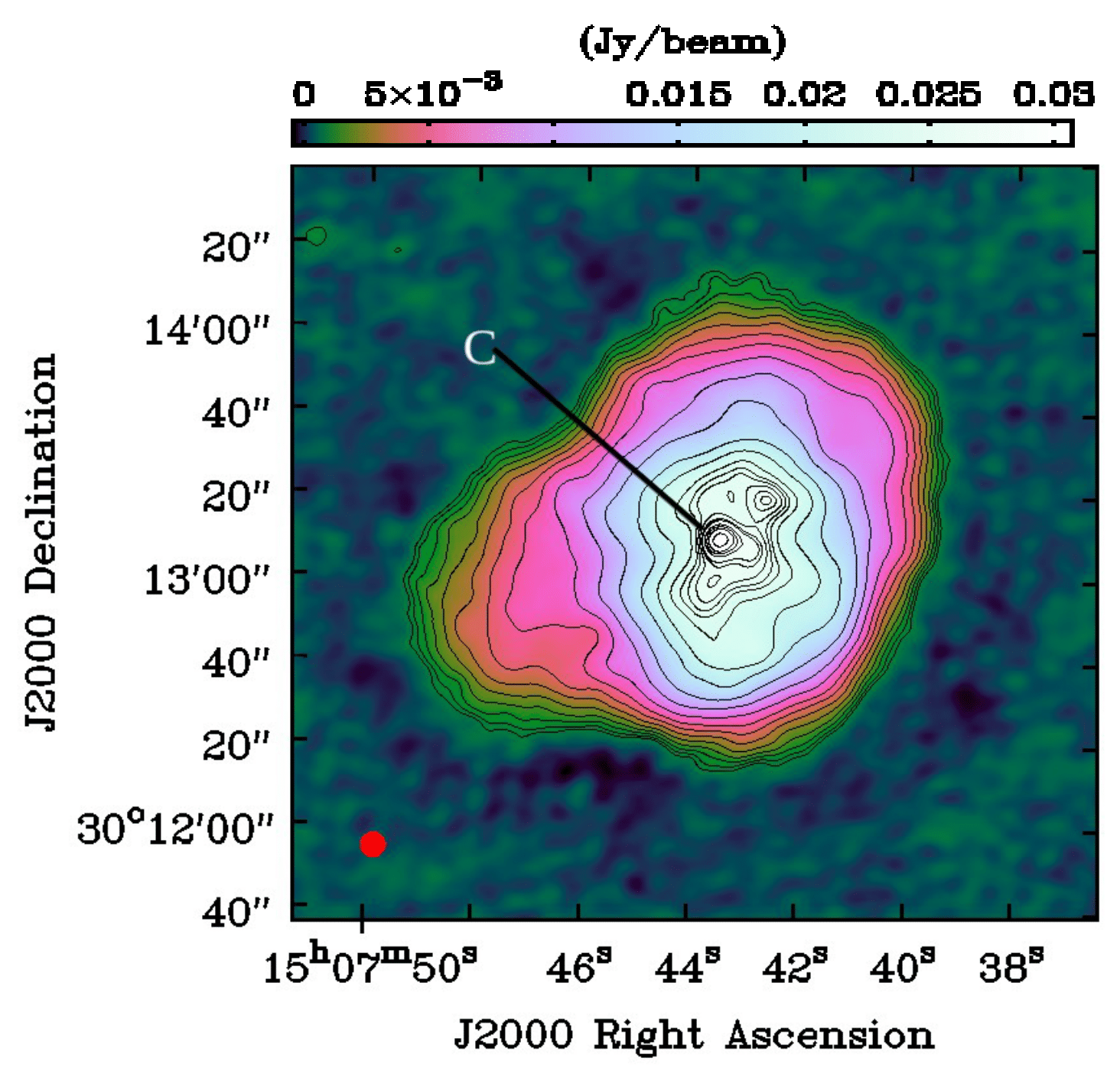}
\caption{LOFAR image of J1507+3013 at 144 MHz (6-arcsec resolution). The contour levels are at 3$\sigma\times$(1, 1.4, 2, 2.8, 4, 5.7, 8, 11, 16, 23, 32, 45, 64, 80, 90, 96, 100, 104, 106, 109, 111, 113, 115, 120, 124, 132), where $\sigma = 77$ $\mu$Jy beam$^{-1}$ is the rms noise in the LOFAR map. The colour scale represents the flux density.}
\label{fig:orc1}
\end{figure}

\subsection{J1507+3013 in other all-sky radio surveys}
\label{subsec:lofar}  
We checked the morphology of J1507 + 3013 in available all-sky radio surveys, such as the Low Frequency Array (LOFAR) Two-meter Sky Survey 2nd data release (LoTSS DR2) at 144 MHz \citep{Sh22}, VLA Sky Survey (VLASS) at 3 GHz \citep{La20}, Westberk Northern Sky Survey (WENSS) at 326 MHz \citep{Re97}, TIFR GMRT Sky Survey (TGSS) at 150 MHz \citep{In17}, NRAO VLA Sky Survey (NVSS) at 1400 MHz \citep{Co98}, GaLactic and Extragalactic All-sky Murchison Widefield Array (GLEAM) at 72--231 MHz \citep{Hu17} in order to confirm the nature
of J1507+3013. Apart from the VLA FIRST survey, the diffused circular-like morphology of J1507+3013 is identified in the LOTSS DR2 and VLASS surveys (other surveys detected J1507+3013 as a point source). We present the LoTSS DR2 image of J1507+3013 at 144 MHz in Fig. \ref{fig:orc1} and the superimposed VLA FIRST survey image of J1507+3013 at 1400 MHz with LoTSS DR2 at 144 MHz in Fig. \ref{fig:orc_2}. 

\subsection{LOFAR view of J1507+3013}
\label{subsec:lofar_view}
J1507+3013 is identified in the FIRST survey at 1400 MHz (see Fig. \ref{fig:ORC_rad_op}) and in LOFAR at 144 MHz (see Fig. \ref{fig:orc1}). The inner structure of the LOFAR image (shown in white and bluish-white in Fig. \ref{fig:orc1}) covers the diffuse emission of J1507+3013, as seen in the FIRST image. We present a superimposed image of FIRST at 1400 MHz in contour with the LOFAR image at 144 MHz in Fig. \ref{fig:orc_2}. The low-frequency map of J1507+3013 at 144 MHz shows extended diffuse emission along the eastern or left-hand side of the structure (denoted by E in Fig. \ref{fig:orc_2}). This diffuse emission of synchrotron plasma may arise because of the interaction of the host galaxy with the inhomogeneous IGM.  

\subsection{Cluster near \src{}}
\label{subsec:cluster} 
We look for available galaxy clusters near J1507+3013 in the Canada France Hawaii Telescope Legacy Survey (CFHTLS) galaxy cluster catalogue \citep{Du11}, the Dark Energy Spectroscopic
Instrument (DESI) survey \citep{Zo21}, and the Two Micron All-Sky Survey (2MASS), Wide-field Infrared Survey Explorer (WISE), SuperCOSMOS clusters of galaxies catalogue \citep{We18}. \src{} is located about 7.7 arcmin from the centre of the WHL J150711.9+300915 cluster. However, J1507+3013 is unlikely to be related to WHL J150711.9+300915 because the cluster is located at a much higher redshift ($z=$0.3363) than J1507+3013 ($z=$0.079). We also checked for any compact group of galaxies around J1507+3013 and found that \src{} is not associated with an over-density of galaxies around it.

\subsection{Optical counterparts of \src{}}
\label{sec:optical_counterparts} 
We look for the optical counterparts of \src{} in Sloan Digital Sky Survey (SDSS), the Panoramic Survey Telescope and Rapid Response System (Pan-STARRS) \citep{Fl20} and the 9th data release from the DESI's Legacy Imaging Surveys \citep[DESI LS DR9;][]{Sc21}.
 A bright optical/infrared source coincides with the radio core (labelled as `C' in Fig. \ref{fig:ORC_rad_op}) and is identified as the optical counterpart of the source that has been seen near the geometrical centre of J1507+3013. The optical counterpart of J1507+3013 is identified from the Pan-STARRS \citep{Fl20}, SDSS, and also with DESI. The source has been previously identified as SDSS J150744.08+301314.4, WISEA J150743.54+301307.8, GALEXASC J150743.38+301308.1, 2MASX J15074353+3013084 and 2MASS J15074352+3013086. A very strong radio emission can be seen at the core of J1507+3013. In Fig. \ref{fig:ORC_rad_op}, we present the VLA FIRST image of J1507+3013 at 1400 MHz in contour with the optical image obtained from Pan-STARRS \citep{Fl20} superimposed. The lowest contour level is maintained at 3$\sigma$, where $\sigma$ is the RMS noise of the source in the FIRST map.

\begin{figure}
\includegraphics[width=9.1cm, origin=c]{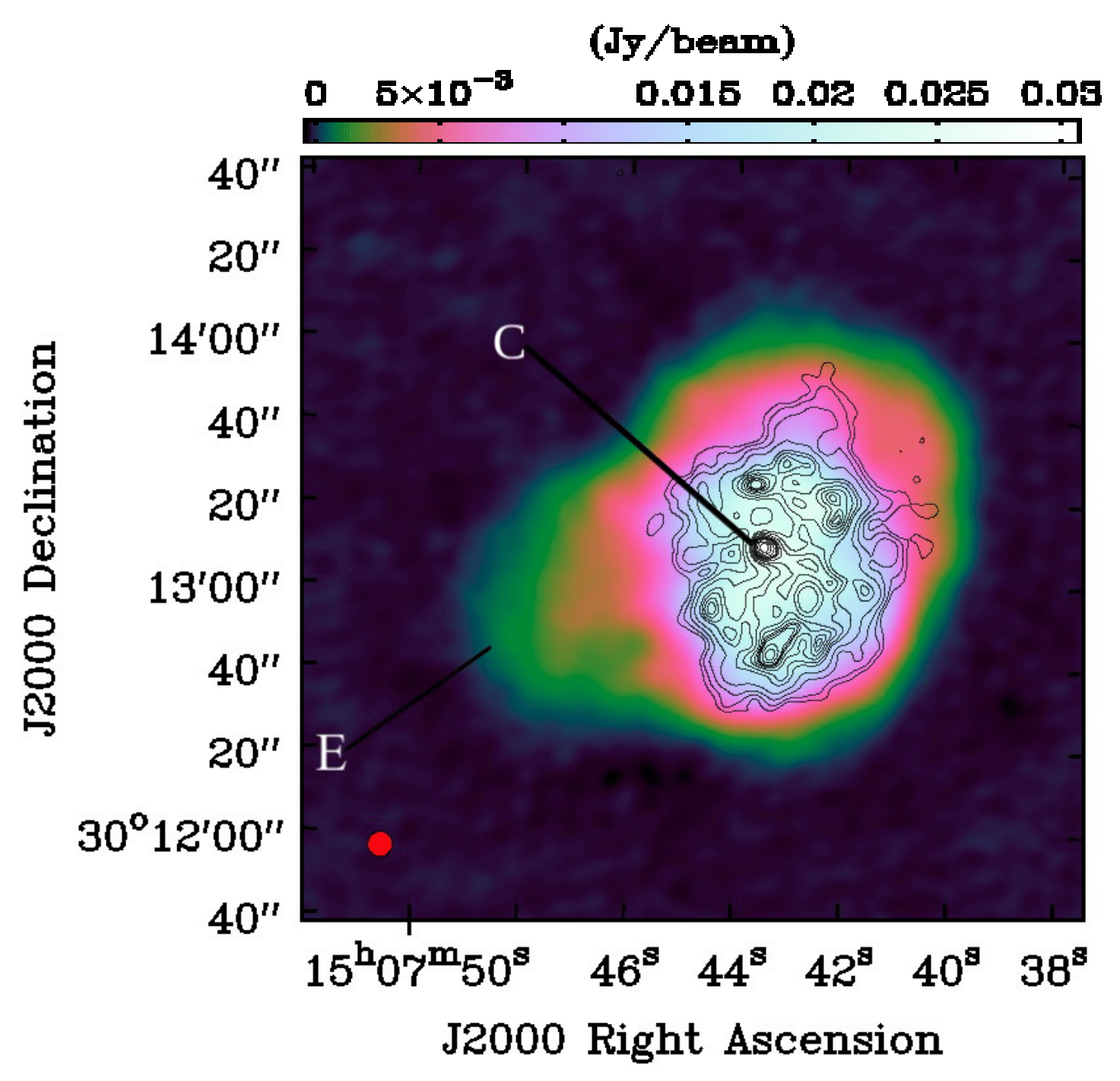}
\caption{VLA FIRST image of J1507+3013 at 1400 MHz in contour [levels: 3$\sigma\times$(1, 1.4, 2, 2.8, 4, 4.5, 5, 5.4, 5.7, 5.8, 5.9, 6, 6.2, 6.5, 7, 7.2, 7.4), where $\sigma=133$  $\mu$Jy beam$^{-1}$ is the rms noise of the FIRST map], with LOFAR at 144 MHz superimposed. The colour scale represents the flux density.}
\label{fig:orc_2}
\end{figure}

\begin{figure*}
\vbox{
\centerline{
\includegraphics[width=9.1cm, origin=c]{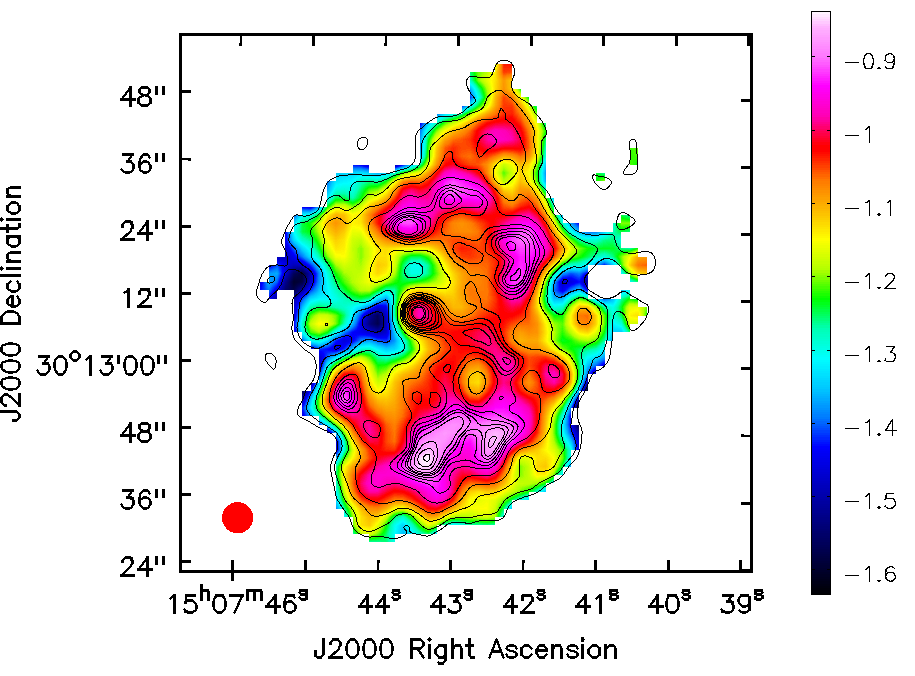}
\includegraphics[width=9.1cm, origin=c]{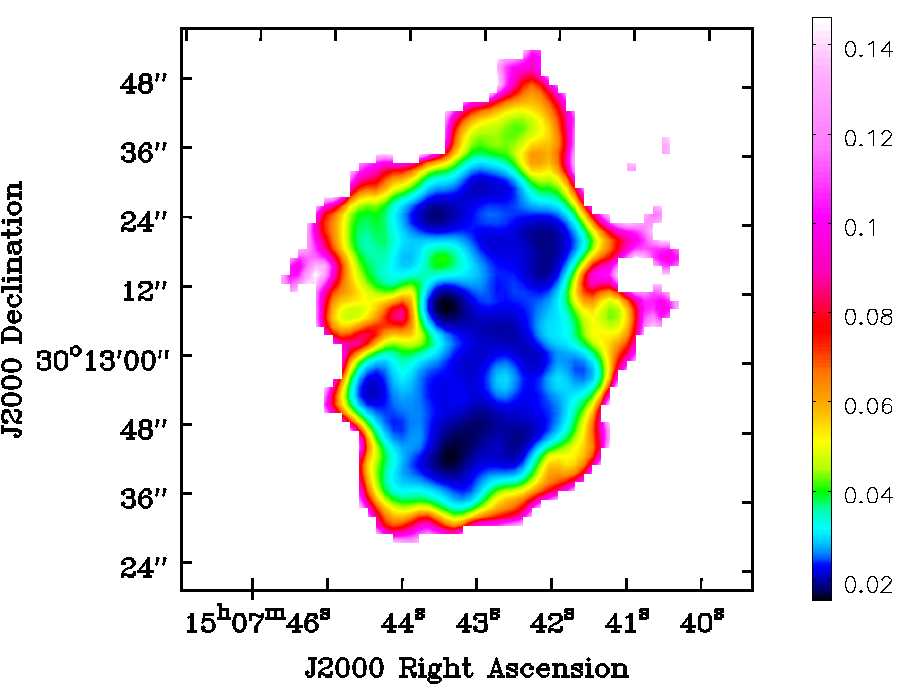}
}
}
\caption{Left: Spectral index map of J1507+3013 between 1400 MHz and 144 MHz with the FIRST image in contour at 1400 MHz superimposed. The contour levels are at 3$\sigma\times$(1, 1.4, 2, 2.8, 4, 4.5, 5, 5.4, 5.7, 5.8, 5.9, 6, 6.2, 6.5, 7, 7.2, 7.4), where $\sigma=133$  $\mu$Jy beam$^{-1}$ is the rms noise of the FIRST map. The colour scale represents the spectral index. Right: Spectral index error map of J1507+3013 between 1400 MHz and 144 MHz. The colour scale represents the error in the spectral index.}
\label{fig:spec_map}
\end{figure*}

\section{Properties of J1507+3013} 
\label{sec:properties} 

\subsection{Spectral properties of J1507+3013}
\label{sec:spec-index-map} 
We present the spectral index map of J1507+3013 between 1400 MHz and 144 MHz and the corresponding error map in Fig. \ref{fig:spec_map}. The typical error in the spectral index map is $\sim0.02$ (see Fig. \ref{fig:spec_map}). The error is more or less uniform, except at the edge of the radio emission. The contour plot in Fig. \ref{fig:spec_map} shows the VLA FIRST image of J1507+3013 at 1400 MHz. As can be seen in Fig. \ref{fig:spec_map}, the spectral index, $\alpha_{144}^{1400}$, near the core of \src{} is approximately --1.0, whereas the spectral index lies in the range of --0.84 to --0.97 at the location of hotspots. The steep spectral index (<--1.0) can be seen along the diffuse emission of \src{}. The spectral steepening in the diffuse emission of \src{} may be due to spectral ageing. J1507+3013 is detected at 74, 144, 150, 408, 1400, 4830, and 4850 MHz. We use flux densities of J1507+3013 in the frequency range 74 MHz--4850 MHz from the published available literature for the study of the spectral energy distribution (SED) at radio wavelengths. SED spectra typically represent radiation laws, along with their parameters such as the power-law energy index or emissivity index. 
The radio SED of galaxies can be separated into two primary domains: the non-thermal domain at $\nu \lesssim 10\,\mathrm{GHz}$ and the thermal domain at frequencies of $10\mbox{--}20\,\mathrm{GHz}<\nu <100$ GHz. We used the non-thermal radio domain to plot the SED for \src{}, as shown in Fig. \ref{fig:SED}.
J1507+3013 exhibits a straight power-low radio spectrum with a best-fitting energy index (spectral index) of --0.90$\pm$0.04 (see Fig. \ref{fig:SED}). The SED from the source has been widely attributed to non-thermal synchrotron radiation coming from relativistic charged particles accelerating in a magnetic field.

\subsection{Radio properties of J1507+3013}
\label{sec:radio} 
\subsubsection{Flux density measurements of J1507+3013}
\label{sec:flux} 
The integrated radio flux densities of J1507+3013 at 1400 MHz (NVSS) and 144 MHz are measured as 350$\pm17$ mJy and 2310$\pm30$ mJy, respectively.
Instead of using the flux density from the VLA FIRST survey, we used NVSS \citep{Co98} flux density at 1400 MHz. The FIRST survey is prone to flux density loss owing to the shortage of antennas at short spacing. The errors associated with our measured flux densities are calculated using the equation \citep{phoenix} 

\begin{equation}
\sigma_I=I \sqrt{2.5 \frac{\sigma^2}{I^2} + 0.05^2},
\label{eq:error}
\end{equation}
where $\sigma_I$ is the total error on the integrated flux density, $\sigma$ is the rms error in the image, and ${\it I}$ is the total integrated flux density of the source.  
We assume 0.05 (5\%) as the instrumental and pointing errors of VLA (as described in \citet{Middelberg08}). 

\subsubsection{Radio luminosity measurements of J1507+3013}
\label{subsec:radio_luminosity} 
For the radio luminosity measurement, we used the following standard formula \citep{Do09}:
\begin{equation}
    L_{\textrm{rad}}=\frac{4\pi{D^{2}_{L}}(z)}{(1+z)^{1-\alpha}}\times S_{1400 ~\text{MHz}}
\end{equation}
where $z$ is the redshift of J1507+3013 ($z=0.079\pm0.004$), $\alpha$ is the spectral index, $D_{L}(z)=1148\times 10^{22}$ m is the luminosity distance to the source in metres (m), $S_{1400~\text{MHz}}$ is the flux density in W m$^{-2}$ Hz$^{-1}$ of J1507+3013 at 1400 MHz. Using the spectral index of --0.90, the radio luminosity of J1507+3013 is measured as 5.0$\times$10$^{24}$ W Hz$^{-1}$ at 1400 MHz.

\subsection{Optical/IR properties of J1507+3013}
\label{sec:Optical} 
The optical host galaxy is an extended elliptical galaxy with a major axis of 26.80 arcsec, minor axis of 21.48 arcsec, and a position angle of 53$^{\circ}$ \citep{Ad08}. The physical extent of the optical host galaxy of \src{}, taking the major axis as the diameter of the elliptical optical galaxy, is 41.8 kpc. The optical galaxy has a photometric redshift of 0.079$\pm$0.004 \citep{De19}. 

As discussed in Subsection \ref{subsec:radio_luminosity}, the radio luminosity of J1507+3013 is 5.0$\times$10$^{24}$ W Hz$^{-1}$ ($\sim$10$^{24.70}$ W Hz$^{-1}$) at 1400 MHz. Such a high luminosity suggests that an AGN with a SMBH resides at the core of J1507+3013.
After examining the relationship between galaxy luminosity, black hole mass, and radio power in nearby active galaxies, \citet{Fr98b} concluded that the radio power of an AGN is a reliable indicator of SMBHs and a good way to estimate their mass. The local mass density of the black hole of J1507+3013 can be estimated using the conversion factors listed below in accordance with the concepts of \citet{Fr98a}:

\begin{equation}
\log_{10}M_{BH} (M_{\odot})=0.376\log_{10}L_{rad}+0.173
\end{equation}
where $L_{rad}$ is the radio luminosity of J1507+3013 at 1400 MHz and $M_{\odot}$ is the solar mass. The above relation gives the mass density of the black hole of J1507+3013 as $\log_{10}$M$_{BH}$ (M$_{\odot})$ = 9.44.

The WISE \citep{Wr10} is an all-sky infrared survey observed in the W1, W2, W3, and W4 bands, which correspond to 3.4, 4.6, 12, and 22-$\mu$m wavelengths. The WISE colours of \src{} are W1 = 12.86, W2 = 12.814, W3 = 12.218, and W4 = 10.63. So W1--W2 = 0.046 and W2--W3 = 0.6, which suggests that the source is possibly a low-excitation radio galaxy (LERG) (W1--W2 < 0.5 and 0.2 < W2--W3 < 4.5) \citep{Pr18, Da20}.

\section{Discussions}
\label{sec:result}
Here, we discuss some possible formation scenarios for the origin of such diffuse emissions, as observed for J1507+3013.
\begin{figure}
\includegraphics[width=8.7cm, origin=c]{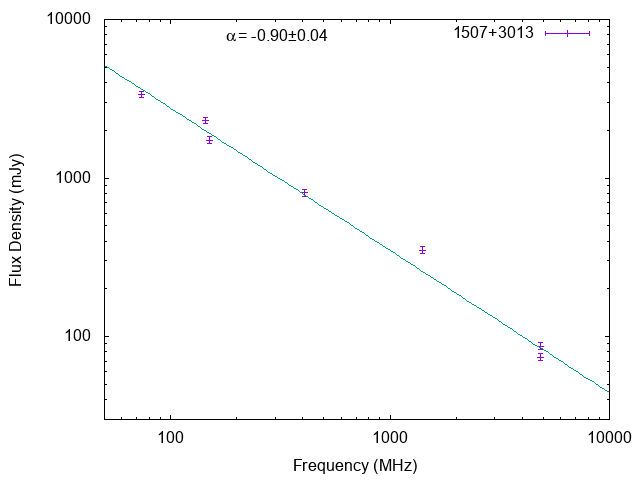}
\caption{Spectral energy distribution (SED) of J1507+3013.}
\label{fig:SED}
\end{figure}

\subsection{Is \src{} an early phase of an ORC?}
\label{subsec:comparison}
\begin{figure}
\includegraphics[width=9.1cm, origin=c]{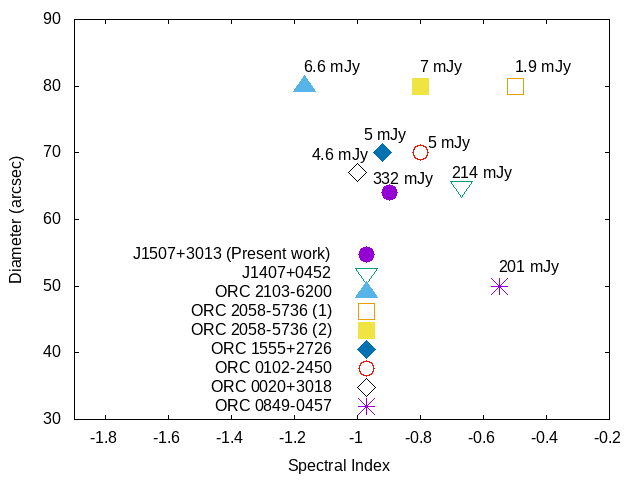}
\caption{Distribution of diameter with spectral index for all previously identified ORCs, including J1507+3013, the source presented in this paper. This plot also includes J1407+0453, a source with a horseshoe-shaped inner ring of diffuse emission \citep{Ku23}. The flux densities at $\sim$1 GHz for each source are noted in the plot. Here filled purple circle with flux density 332 mJy represents J1507+3013.}
\label{fig:optical2}
\end{figure}

The morphology and spectral behaviour of J1507+3013 are similar to those of other discovered ORCs \citep{No21a, No21b, No21c, Ko21, Om22, No22}. Figure \ref{fig:optical2} shows the distribution of the spectral index with the diameter for the previously discovered ORCs, along with the source (J1507+3013) presented in this paper (with a purple-filled circle). This plot also includes J1407+0453, a recently identified source with a horseshoe-shaped ring (HSR) of diffuse emission \citep{Ku23}. We have added the flux density of each source measured at 1 GHz to the plot. From this plot, it can be seen that the spectral indices of six previously identified ORCs, one candidate ORC (J084927.3--045732.3) \citep{Gu22}, J1407+0453 and the source presented in this paper are in the range of --0.55 to --1.19. As can be seen in Fig. \ref{fig:optical2}, all ORCs have a flux density $\leq$7 mJy at $\sim$ 1 GHz, except for the source presented in this paper (with a flux density of 332 mJy at $\sim$ 1 GHz, using $\alpha=-0.90$, source J1407+0453 presented by \citet{Ku23} (with a flux density of 214 mJy at $\sim$ 1 GHz, using $\alpha=-0.67$), and the candidate ORC J084927.3--045732.3 presented by \citet{Gu22} (with a flux density of 201 mJy at $\sim$ 1 GHz). As shown in Fig. \ref{fig:optical2}, the diameters of all sources are in the range of 50--95 arcsec. The peak-to-peak (H1-to-H4) distance of J1507+3013 (see Fig. \ref{fig:optical2}) at 1400 MHz (FIRST image; see Figs. \ref{fig:ORC_rad} and \ref{fig:ORC_rad_op}) is 44 arcsec with a physical extent of approximately $\sim$68 kpc. For comparison with the ORCs in Fig. \ref{fig:optical2}, we used the largest angular size (LAS) of \src{}, namely 64 arcsec. The physical extent of J1507+3013 in the LOFAR image at 144 MHz extends up to $\sim$183 kpc with a 2.2 arcminute diameter because of the elongated diffuse emission (labelled as E in Fig. \ref{fig:orc_2}) in the LOFAR image (not included in Fig. \ref{fig:optical2}).

Previously, only three ORCs (ORC1, ORC4, and ORC5) were found with optical host galaxies near their geometrical centres. If J1507+3013 is an ORC (owing to its similar morphology), it would be the fourth source with an optical core.
The range of photometric redshifts for the previously identified ORCs is between 0.27 to 0.55. So, \src{} would be the closest source discovered so far in this category. The measured radio luminosities for ORC1, ORC4, and ORC5 are $\sim$10$^{23}$ W H$z^{-1}$, whereas that of \src{} is $\sim$10$^{24.70}$ W H$z^{-1}$. The blackhole mass for ORC5 is 7.5$\times10^8$ M$_{\odot}$, whereas for \src{} the black hole mass is calculated as $\sim$2.8$\times10^9$ M$_{\odot}$. Therefore, even with the similarity of morphology and spectral properties, \src{} is more luminous and massive than previously discovered ORCs. However, only three ORCs have been detected so far with identified optical cores; we need more samples and multi-wavelength observations to compare the behaviour of the central galaxy of \src{} with ORCs.

\subsection{Is \src{} a radio remnant giant blast wave?}
\label{subsec:remnant}
Earlier, it was proposed that circularly symmetry-diffused sources (such as ORCs) containing one or more optical galaxies near the geometrical centre of the structure are the result of a giant blast wave at the core of the optical host galaxy \citep{No21a, No21b, No21c}, which results in a ring or spherical shell of radio emission. Similar to SNRs and planetary nebulae, such spherical shells would appear as edge-brightened discs \citep{No21a, No21b, No21c}. This radio emission is likely a synchrotron emission caused by electrons accelerated by a shock \citep{Do02, Ca12, Ri21}. 
A binary SMBH merger in the centre of a galaxy can result in a spherical shock \citep{Bo12}. In this instance, we anticipate that the magnetic field in the ring is largely tangential, orthogonal to the velocity of the shock, and resembles that of SNRs or cluster relics. 

J1507+3013 shows multiple hotspots (strong radio components) in the inner structure, denoted by H1, H2, H3, H4, and H5, with an optical/IR counterpart coinciding with the radio core denoted by C. The measured flux densities of the radio emissions at the denoted hotspots are 12 mJy (H1), 9 mJy (H2), 4 mJy (H3), 10 mJy (H4), and 2 mJy (H5); the radio core has a flux density of 6 mJy. The multiple hotspots of J1507+3013 and the steepening of these hotspots (--0.85 to --0.95) in the spectral index map (see Fig. \ref{fig:spec_map}) provide significant evidence of the electron population that could be affected by the ageing of the synchrotron spectrum owing to energy losses. This population of electrons is possibly circling after strong interaction with the ISM (evidence of multiple hotspots), which causes the circularly-symmetric structure of J1507+3013. The hotspots of the strong radio counterparts are possibly radio remnants formed because of the giant blast wave of the central galaxy. Recently, \citet{Do23} provided a simulation study in which they observed the generation of such diffuse emission structures, as seen in \src{}. \citet{Do23} used advanced simulations to demonstrate that diffuse circular emission as observed in \src{} could emerge naturally as a result of shocks produced by extreme galactic merger events which result in a galaxy with a virial mass of $\sim$10$^{13} $M$_\odot$. More samples of such diffuse sources hosted by optical galaxies are required to confirm the above possibility.

\subsection{Is \src{} a mini halo?}
\label{subsec:halo}
Usually, mini-halos are located in the core of cooling flow clusters. Their sizes vary from a few hundred kiloparsecs to half a megaparsec. Mini-haloes, defined as regions of diffuse emission with a steep spectral index and low surface brightness, are present around powerful radio galaxies at the cores of certain clusters, such as PKS 0745--191 \citep{Ba91}, Perseus \citep{Si93}, Virgo \citep{Ow00}, Phoenix cluster \citep{va14}, RXCJ1720.1+2638 \citep{Gi14} and recently the Abell 1413 cluster \citep{Ri23}.

As discussed in the Subsection \ref{subsec:cluster}, \src{} is located about 7.7 arcmin (with a projected distance of 2.4 Mpc) from the centre of a cluster (WHL J150711.9+300915) with a redshift of 0.3363, which is unlikely to be responsible for the formation of J1507+3013 because of its large distance and lack of redshift similarity to J1507+3013. Based on the previous study, no cluster halo exhibits circularly symmetric emission, like the diffuse emission seen in J1507+3013; instead, cluster halos are observed as patchy radio emissions, and a cluster of galaxies is typically present together with a cluster halo emission \citep{No21b}. Therefore, we can conclude that J1507+3013 is probably not a cluster halo, because no cluster of galaxies is found near it. On the other hand, previously discovered ORCs appear to be associated with either an overdensity or a nearby companion \citep{No21c, No22}.

\subsection{Is \src{} a fossil radio galaxy?}
\label{subsec:fossils}
A fossil radio galaxy displays radio emissions that reveal evidence of prior AGN activity. These objects often exhibit extended radio structures such as lobes or jets, which are remnants of the prior activity of a central AGN. As seen in previous studies, fossil radio galaxies should have a very steep spectral index (<--1.5) \citep{Kem04, Ri23}.
The diffuse emission of \src{} may represent the remains of a fossil radio galaxy; however, as seen in the spectral index map of \src{}, the spectral index lies in the range of --0.9 to --1.4 (see Fig. \ref{fig:spec_map}). The integrated fitted spectral index is also noted as --0.9, as shown in Fig. \ref{fig:SED}. Therefore, \src{} may not be a fossil radio galaxy. 

\subsection{Is \src{} a conventional radio galaxy?}
\label{subsec:double_RG}
Radio galaxies consist of a SMBH at the centre that ejects beams of intense particles (most commonly electron particles). These intense beams of particles form jet structures pointing in two opposite directions. Radio galaxies are generally categorized into low-power FR-I and high-power FR-II classes \citep{Fa74}. Galaxies in the FR-I class consist of collimated jets and are typically bright near the core of the structure but quickly fade at the edge. Galaxies in the FR-II class, on the other hand, show no systematic variation
in radio emission near the core and have compact hotspots at the outer edge of the structure. The radio power of the FR-I sources is $\leq$2$\times$10$^{25}$ W Hz$^{-1}$, whereas it is $\geq$2$\times$10$^{25}$ W Hz$^{-1}$ for FR-II sources \citep{Fa74}. J1507+3013, with radio power 5.0$\times$10$^{24}$ W Hz$^{-1}$ ($\sim$10$^{24.70}$ W Hz$^{-1}$) at 1400 MHz, shows no evidence of ejection of the particles in two opposite directions from the core, as is normally seen in radio galaxies, and most likely the population of particles (electrons) is circling the optical galaxy of \src{}. For radio galaxies, spectral steepening has been observed from the core to the edge of the structure, and the core typically has a flatter spectral index compared with the outer lobe; however, if we consider H1-C-H4 as a radio galaxy, where H1 and H4 are the outer edges, then H1 and H4 should consist of a steeper spectral index than core (C), but H1 and H4 consist of a flatter spectral index than the core (C). Therefore, based on the current observation, \src{} may not be a conventional radio galaxy; rather, it may be a diffuse emission surrounded by an elliptical galaxy in which particles (electrons) are circling the optical galaxy because of a catastrophic event, such as a merger that occurred in the central galaxy (as discussed in subsection \ref{subsec:remnant}). Further observations are required to better understand the nature of \src{}.

\section{Conclusions}
\label{sec:conclusion}
This paper reports an extended diffuse circularly symmetric radio emission, J1507+3013, from the VLA FIRST survey at 1400 MHz. The source presented in this paper is also identified in the LOFAR at 144 MHz with extended diffuse emission. The physical extent of \src{} is $\sim$68 kpc in the VLA FIRST image, whereas, with extended diffuse emission in the LOFAR image, the physical extent is $\sim$183 kpc (comparable to that of previously discovered ORCs).
J1507+3013 has an optical/IR counterpart with a redshift $z=0.079$ near the geometrical centre of the structure. It has a stronger flux density compared with previously identified circular diffused sources, such as ORCs, and is proximately close ($z = 0.079$). J1507+3013 is luminous maybe because it is in the early phase of evolution, yet to lose its energy and become a low-power faint ORC, similar to other discovered ORCs. We discuss some possible scenarios for the formation of J1507+3013; however, we are still unsure of the mechanism for its formation. More samples of such sources are required to understand the emission mechanisms of this type of source. Further multi-wavelength follow-up observations of this kind of source are encouraged so that the nature of these objects can be investigated.

\section*{acknowledgments}
We thank the anonymous reviewer for helpful suggestions. This publication makes use of data from LOw-Frequency ARray (LOFAR) Two-metre Sky Survey 2nd data release (LoTSS DR2). LOFAR is the Low-Frequency Array designed and constructed by ASTRON. It has observing, data processing, and data storage facilities in several countries, which are owned by various parties (each with their own funding sources), and which are collectively operated by the ILT foundation under a joint scientific policy. The efforts of the LSKSP have benefited from funding from the European Research Council, NOVA, NWO, CNRS-INSU, the SURF Co-operative, the UK Science and Technology Funding Council, and the Jülich Supercomputing Centre. 
This publication also uses Pan-STARRS1 survey data. The Pan-STARRS1 Surveys (PS1) have been made possible through contributions of the Institute for Astronomy, the University of Hawaii, the Pan-STARRS Project Office, the Max-Planck Society and its participating institutes, the Max Planck Institute for Astronomy, Heidelberg and the Max Planck Institute for Extraterrestrial Physics, Garching, The Johns Hopkins University, Durham University, the University of Edinburgh, Queen's University Belfast, the Harvard-Smithsonian Center for Astrophysics, the Las Cumbres Observatory Global Telescope Network Incorporated, the National Central University of Taiwan, the Space Telescope Science Institute, the National Aeronautics and Space Administration under Grant No. NNX08AR22G issued through the Planetary Science Division of the NASA Science Mission Directorate, the National Science Foundation under Grant No. AST-1238877, the University of Maryland, and Eotvos Lorand University (ELTE).

\section*{Data Availability}
We used the data from the catalogue of the VLA FIRST survey \citep{Be95, Wh97}, Pan-STARRS1 \citep{Fl20}. The FIRST Database is publicly available at \href{http://sundog.stsci.edu/index.html}{http://sundog.stsci.edu/index.html}. The Pan-STARRS1 survey data is publicly available at \href{http://ps1images.stsci.edu/cgi-bin/ps1cutouts}{http://ps1images.stsci.edu/cgi-bin/ps1cutouts}. We also used LOw-Frequency ARray (LOFAR) Two-metre Sky Survey 2nd data release (LoTSS DR2) available at \href{https://lofar-surveys.org/}{https://lofar-surveys.org/}. The data that support the figures and plots within this paper and the other findings of this study are available from the corresponding author upon reasonable request.





\label{lastpage}
\end{document}